\documentclass[twocolumn,showpacs,preprintnumbers,amsmath,amssymb,a4paper]{revtex4}
\usepackage{epsfig}

\begin{document}

\title{Efimov effect in nuclear three-body resonance decays}

\author{E. Garrido}
\affiliation{Instituto de Estructura de la Materia, CSIC, Serrano 123, E-28006
Madrid, Spain}
\author{D.V. Fedorov}
\author{A.S. Jensen}
\affiliation{Department of Physics and Astronomy, University of Aarhus, 
DK-8000 Aarhus C, Denmark}

\begin{abstract}
We investigate the effects of the nearly fulfilled Efimov conditions on the properties of three-body resonances. Using the hyper-spheric adiabatic expansion method we compute energy distributions of fragments in a three-body decay of a nuclear resonance.  As a realistic example we investigate the $1^-$ state in the halo nucleus $^{11}$Li within a three-body $^9{\rm Li}+n+n$ model. Characteristic features appear as sharp peaks in the energy distributions.  Their origin, as in the Efimov effect, is in the large two-body $s$-wave scattering lengths between the pairs of fragments.
\end{abstract}


\maketitle

\paragraph*{Introduction.}

The Efimov effect was introduced more than thirty years ago as an
anomaly in a three-body system arising when at least two of the three
two-body $s$-wave scattering lengths approach infinity
\cite{efi70}. Then an increasing number of three-body bound states
appear close to the two-body threshold even if
there are no two-body bound states. The effect is
prohibited by the Coulomb potential while only diminished by higher
angular momentum \cite{nie01}.  Although entirely
possible in molecules \cite{jen04} the effect is unlikely to appear in
nuclei due to unfavorable mass ratio \cite{fed94,brasil}.

Still there exists a number of nuclear systems, called halos \cite{jen04}, which are natural three-body systems -- a core plus two neutrons -- where the Efimov condition of at least two large scattering lengths is nearly fulfilled. Although the unfortunate combination of the heavy core and light neutrons prohibits the appearance of bound Efimov states in the discrete spectrum, they still may appear as peculiar structures in the continuum. Very little, however, is known theoretically about the Efimov states in the continuum.

Experimentally, on the other hand, the number of accurate and kinematically complete experiments for three-body decays of nuclear resonances is rapidly increasing \cite{fyn00,bla03}.  Also three-body decays of excited states of 
small molecules are presently experimentally investigated in details \cite{gal04}.  The measured observables are the width and, particularly promising, the energy distributions of the three fragments after the decay
\cite{fyn03}.  Although a number of theoretical studies involve calculations of three-body continuum properties \cite{glo96,kie01,laz05}, calculations of energy distributions for resonances under Efimov conditions have not been done before.  In contrast to bound states, investigations of the fingerprints of the Efimov effect on the decays of three-body resonances are so far lacking.

In this letter we report on an investigation of nuclear three-body resonances under the nearly fulfilled Efimov conditions. In particular we calculate the energy distributions of the decay fragments of a nuclear resonance and trace the origin of the characteristic peaks in these distributions.

\paragraph*{Calculation of energy distributions.}

We assume that three particles emerge after decay of a preformed
resonance state.  At large distance we then strictly have a three-body
problem.  This is not necessarily true at small distance where the
three-body cluster assumption may be inappropriate. We shall extend
the concept from two-body nuclear $\alpha$-decay.  There the outer part of the potential between the
daughter nucleus and the $\alpha$-particle is known and the inner part
is adjusted to give the correct resonance energy.  This treatment
accounts for the major variations of $\alpha$-decay widths. The
fine-tuning is obtained by using the preformation factor describing
the probability for finding an $\alpha$-particle at the inner turning
point of the two-body potential.  For two-body
decays the width is determined by the outer part of the wave-function, while the fragment energy is fixed from energy conservation.
It is anticipated and intuitively plausible that for three-body
decays both the width and the fragment energy distributions are determined by the large distance behavior of the wave-function.

The notion of large distance is not a priori obvious for three particles where either all three or only two inter-particle distances can be large.  We shall specify distances by the value of the hyper-radius $\rho$,
\begin{equation}
m\rho^2 = \sum_{i=1}^{3}m_ir_i^2\;,
\end{equation}
where $m_i$ is the mass and ${\bf r}_i$ the c.m. coordinate of the particle number $i$, and $m$ is an arbitrary mass scale. The other five hyper-spheric coordinates are dimensionless angles, $\Omega$, which determine the directions and relative values of the coordinates of the constituents \cite{nie01}.

Within the hyper-spheric adiabatic method one distinguishes the fast, $\Omega$, and slow, $\rho$, coordinates. Then, for every fixed slow coordinate $\rho$ the eigen-value problem is solved for the fast angular coordinates $\Omega$
\begin{equation}\label{hrho}
H(\rho)\Phi_n(\rho,\Omega) = W_n(\rho)\Phi_n(\rho,\Omega)\;,
\end{equation}
where $H(\rho)$ is is the Hamiltonian of the three body system with fixed $\rho$, $W_n(\rho)$ are the angular eigen-values, and $\Phi_n(\rho,\Omega)$ is the angular eigen-function with a parametric dependence on $\rho$. In practice instead of the Schr\"{o}dinger equation (\ref{hrho}) we solve the mathematically equivalent Faddeev equations \cite{complex}.

Then $\Phi_n(\rho,\Omega)$ are used as a basis for the total wave-function $\Psi(\rho,\Omega)$,
\begin{equation}
\Psi(\rho,\Omega) =
\rho^{-5/2} \sum_{n=1}^{\infty} f_n(\rho) \Phi_n(\rho,\Omega)\;,
\end{equation}
where the expansion coefficients $f_n(\rho)$ satisfy the system of hyper-radial equations where the eigen-values $W_n(\rho)$ of the fast angular subsystem serve as effective potentials,
\begin{eqnarray}\label{hr}
&&\left( -\frac{\partial^2}{\partial\rho^2} + \frac{15/4}{\rho^2} +
\frac{2m}{\hbar^2}(W_n(\rho) - E) \right)f_n(\rho) \nonumber \\
&&= \sum_{n'=1}^{\infty}\hat{P}_{nn'}f_{n'}(\rho)\;,
\end{eqnarray}
where $E$ is the three-body energy and $\hat{P}_{nn'}$ are the non-adiabatic terms \cite{nie01}.

A three-body resonance corresponds to a (complex-energy) solution $\Psi(\rho,\Omega)$ of the system (\ref{hr}) with the asymptotic boundary condition of an outgoing wave in every channel $n$,
\begin{equation}\label{exp}
f_n(\rho\rightarrow\infty) = C_n \exp(+i\kappa\rho)\;,
\end{equation}
where $C_n$ is an asymptotic normalisation coefficient and $\kappa=\sqrt{2mE/\hbar^2}$ is the three-body momentum, which is the conjugate of $\rho$. The other five variables in momentum space are the momentum angles $\Omega_\kappa$ which are conjugates of the coordinate space angle $\Omega$. The angles $\Omega_\kappa$ determine the directions and relative values of the momenta of the fragments. In practice instead of the exponential function in (\ref{exp}) we use a corresponding Hankel function. We also employ the so called complex scaling \cite{complex} with an angle $\theta$, $\rho\rightarrow e^{\theta}\rho$, which improves the numerical accuracy of the calculations.

\begin{figure}
\begin{center}
\centerline{\psfig{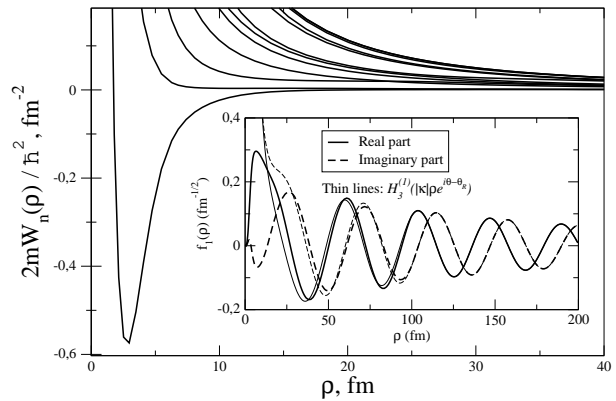}}
\end{center}
\vspace*{-0.8cm}
\caption{The lowest adiabatic potentials $W_n(\rho)$ for the $^{11}$Li(1$^{-}$) halo nucleus within the three-body $^9$Li+n+n model with interactions from \cite{gar02}. The $n$-core and $n$-$n$ scattering lengths are $a_{nc}\approx a_{nn}\approx$20~fm. The inset shows the lowest hyper-radial resonance function with its large distance asymptotics -- the Hankel function. The complex scaling angle $\theta=0.15$, the resonance angle $\theta_R=0.12$ corresponds to the resonance energy of about $0.4-0.1i$~MeV.}
\label{fig-l}
\end{figure}

An example of the effective potentials and the hyper-radial resonance function is shown on Fig.~\ref{fig-l}. The lowest potential has an attractive pocket where a resonance resides. The corresponding dominating component of the hyper-radial function shows a typical resonance behavior -- a pronounced maximum in the pocket region with subsequent oscillations, matching the corresponding Hankel function, in the asymptotic region.

As shown in \cite{fed04} the momentum distributions $P(\Omega_\kappa)$ of the fragments can be calculated from the wave-function $\Psi(\rho,\Omega)$ as
\begin{equation}\label{po}
P(\Omega_\kappa) = | \sum_{n=1}^{\infty}
C_n\Phi_n(\rho\rightarrow\infty,\Omega=\Omega_\kappa) |^2\;.
\end{equation}
For the graphical presentation we shall integrate out the four directional angles in $\Omega_\kappa$ and only leave one angle which specifies the energy $E_i$ of the fragment $i$ relative to the maximum possible energy $E_i^{\rm (max)}$ of this fragment in the decay process.

Different decay mechanisms carry distinct signatures of the process,
e.g. sequential decay via a two-body resonance must give a peak in the
energy distribution of the third particle at the peak-energy preferred
by the first two-body decay. However, when the two-body $s$-wave interaction is
only weakly attractive -- no bound states exist -- no stable intermediate structures
are available for such a sequential decay.  Yet some characteristic effects might appear
especially when the threshold for binding is approached for more than
one two-body $s$-state, i.e. when the Efimov conditions are nearly fulfilled.

\paragraph*{Investigated systems.}

The nuclear candidates for the Efimov conditions are systems of two neutrons and an ordinary core-nucleus where as needed only short-range interactions are present. These structures are found as ground or excited two-neutron halo states at the neutron drip-line or for more stable nuclei \cite{jen04}.

We take the (1$^-$) state in the $^{11}$Li halo nucleus within the three-body $^9$Li+n+n model as a realistic example. The interactions from \cite{gar02} provide the $n$-core and $n$-$n$ scattering lengths of about 20~fm and the effective ranges of about 5~fm.

The adiabatic potentials $W_n(\rho)$ for this system are shown on Fig.~\ref{fig-l}.  The lowest potential shows an attractive pocket with a relatively long tail, extending up to about 20~fm, that is $\rho\leq a_{nc}$, despite the fact that the binary interactions are short-range Gaussians with the range of about 5~fm. This long tail is due to interference between the subsystems with long $s$-wave scattering lengths \cite{nie01}. This tail would lead to the bound Efimov states if only the mass ratio of the constituents were inverse, or if the scattering lengths were larger.

We solve the hyper-radial equations (\ref{hr}) with the boundary condition (\ref{exp}) and obtain the resonance hyper-radial wave-function. The dominating lowest component is shown in the inset of Fig.~\ref{fig-l}.

From the hyper-radial wave-function we extract the asymptotic normalization coefficients in (\ref{exp}) and calculate the energy distributions of the core, $^9Li$, and the neutrons using the expression (\ref{po}).

In the realistic calculation apart from the large scattering length $s$-wave interactions also the interactions in higher partial waves, including spin-orbit and tensor forces are included. Those interactions might affect the distributions and mask the fingerprints of the Efimov conditions. We therefore in addition present two model calculations -- one with only $s$-wave neutron-core interactions, and one where also the $s$-wave neutron-neutron interaction is included. Thus only effects of the large scattering lengths can be seen in the distributions.

For the model systems we restore the resonance energy by increasing the scattering length by a factor of two ($a_{nc}\approx a_{nn}\approx 50$~fm) and by tuning the three-body force. The latter is short-range and therefore  does not affect the asymptotics of the system. The larger scattering lengths also serve the purpose of amplifying the visible signatures of the Efimov conditions.

\paragraph*{Convergence.}
The limit $\rho\rightarrow\infty$ in (\ref{po}) is reached in the region where the tail of the lowest adiabatic potential becomes negligible, that is somewhere about and beyond $a_{nc}$. In practice we calculate the distributions at some finite $\rho_{\rm max}\approx a_{nc}$ and then check that an increase in $\rho_{\rm max}$ does not alter the distributions.

In the hyper-spheric adiabatic method the increase in $\rho_{\rm max}$ demands a corresponding increase in the number $N$ of the adiabatic potentials which have to be included in the wave-function. Thus the convergence has to be checked for the two parameters, $\rho_{\rm max}$ and $N$.

In our numerical convergence tests we calculated the distributions with parameters $3\le N\le 12$ and $a_{nc}\lessapprox\rho_{max}\lessapprox 2a_{nc}$ and checked that there is a region in the parameter space, where the distributions are unchanged.

The model systems converge much easier and demand smaller $\rho_{max}$ and $N$.

\begin{figure}
\centerline{\psfig{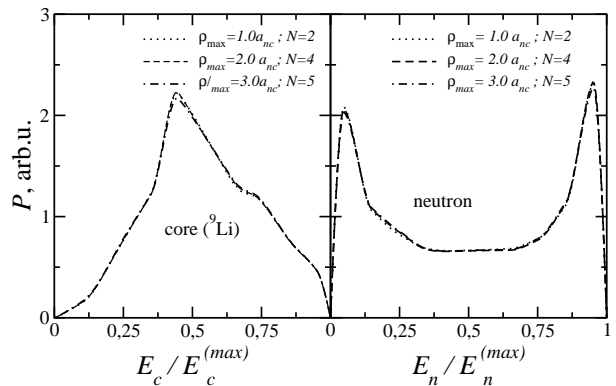}}
\vspace*{-0.4cm}
\caption{The energy distributions of the fragments -- the core, $^9$Li, and the neutrons -- in the decay of a three-body resonance $^{11}$Li$(1^-)$ calculated in the three-body $^9{\rm Li}+n+n$ model with only $s$-wave $n$-core interactions (scattering length $a_{nc}\approx 50$~fm). The different curves are calculated with different $\rho_{max}$ and different numbers of adiabatic channels $N$ to illustrate the convergence.}
\label{fig-m1}
\end{figure}

\paragraph*{Numerical results.}
Let us first look at the model system with only $s$-wave $n$-core interactions. The energy distributions of fragments are shown on Fig.~\ref{fig-m1} where $E$ is the fragment's energy and $E^{\rm (max)}$ is the largest possible energy. The convergence for this system is reached fairly well already at $\rho_{max}=a_{nc}$ and $N=2$.

The core distribution shows one sharp peak at $E_c/E_c^{\rm(max)}\approx 0.5$ while the neutron distribution shows two sharp peaks, at $E_n/E_n^{\rm(max)}\approx 0$ and $E_n/E_n^{\rm (max)}\approx 1$. Since only $s$-wave interactions are allowed these peaks must be the sought fingerprints of the large scattering lengths or, in other words, virtual states.

The geometric interpretation of these peaks is that the decay proceeds by emission of one neutron with maximum energy,
\begin{equation}
E_n^\mathrm{(max)}=E_r \frac{m_c+m_n}{m_c+m_n+m_n} \;,
\end{equation}
(where $E_r$ is the energy of the resonance, $m_c$ and $m_n$ are the core and the neutron masses) allowing the exploitation of the other neutron-core virtual state with vanishing relative energy. This produces the neutron peak close to $E_n/E_n^\mathrm{(max)}=1$.

The other neutron and the core together pick the rest of the energy, $E_r - E_n^\mathrm{(max)} = E_r\frac{m_n}{m_c+m_n+m_n}$, and share it proportional to their masses, since they have vanishing relative energy and thus the same velocity. Consequently the core gets $E_c=E_r\frac{m_n}{m_c+m_n+m_n}\frac{m_c}{m_c+m_n}$ and the other neutron gets $E_{n}=E_r\frac{m_n}{m_c+m_n+m_n}\frac{m_n}{m_n+m_c}$.

Since $m_n\ll m_c$ we have for the predominant core energy
\begin{equation}
E_c = E_r\frac{m_n}{m_c+m_n+m_n}\frac{m_c}{m_c+m_n}
\approx \frac{1}{2}E_c^\mathrm{(max)} \;,
\end{equation}
where
\begin{equation}
E_c^\mathrm{(max)}=E_r \frac{m_n+m_n}{m_c+m_n+m_n}\;.
\end{equation}

Thus the peak in the core distribution should be, in agreement with calculation, just below $E_c/E_c^\mathrm{(max)}=0.5$.

The other neutron has the predominant energy
\begin{equation}
E_{n}=E_r\frac{m_n}{m_c+m_n+m_n}\frac{m_n}{m_n+m_c}\ll E_n^\mathrm{(max)}
\end{equation}
which results, again in agreement with the calculation, in a second peak around $E_n/E_n^\mathrm{(max)}\approx 0$.

The neutron-neutron $s$-wave interaction will allow an additional configuration, the two neutrons with vanishing relative energy against the core with maximum energy. Consequently, another peak should appear in the $E_c/E_c^\mathrm{(max)}$ distribution close to $E_c/E_c^\mathrm{(max)}\approx 1$.

The two neutrons get the rest of the energy, $E-E_c^\mathrm{(max)}=E_r\frac{mc}{m_c+m_n+m_n}$ which they share equally. Thus a neutron will predominantly have the energy
\begin{equation}
E_n = \frac{1}{2}E_r\frac{m_c}{m_c+m_n+m_n}\approx\frac{1}{2}E_n^\mathrm{(max)},
\end{equation}
which again should result in an additional peak just below $E_n/E_n^\mathrm{(max)}\approx\frac{1}{2}$.

\begin{figure}
\begin{center}
\centerline{\psfig{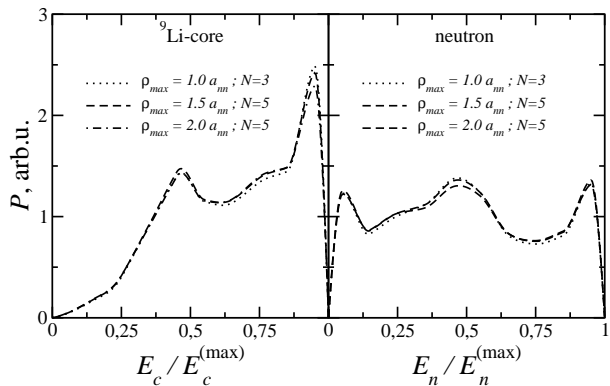}}
\end{center}
\vspace*{-0.9cm}
\caption{The same as fig.~\ref{fig-m1} with additional $s$-wave interaction in the $n-n$ subsystem ($a_{nn} \approx a_{nc}\approx 50$~fm).}
\label{fig-m2}
\end{figure}

\begin{figure}
\begin{center}
\centerline{\psfig{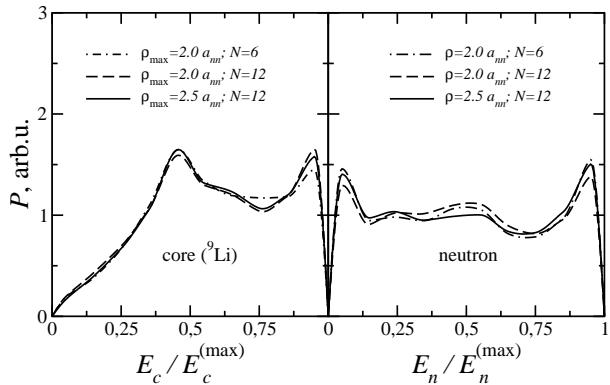}}
\end{center}
\vspace*{-0.8cm}
\caption{The energy distributions of the fragments -- the core, $^9Li$, and the neutrons -- in the decay of a three-body resonance $^{11}$Li$(1^-)$ calculated with interactions from \cite{gar02} where the $n$-core and $n$-$n$ scattering lengths are about $20$fm.}
\label{fig-r}
\end{figure}

These additional structures are clearly seen in Fig.~\ref{fig-m2} where the corresponding numerical distributions are shown. Indeed there are two additional peaks at $E_c/E_c^\mathrm{(max)}\approx 1$ and $E_n/E_n^\mathrm{(max)}\approx\frac{1}{2}$. Convergence is somewhat harder to achieve compared to the first model system and one can see that for $\rho_{max}=2a_{nc}$ five adiabatic channels are not quite enough.

If the virtual states are the dominating properties of the system, the full realistic calculation should exhibit structures, similar to Fig.~\ref{fig-m2}. And so it does, as seen on Fig.~\ref{fig-r} where the realistic distributions are shown. Despite the peaks being somewhat blurred due to interference with higher partial waves, the characteristic structures at $E_c/E_c^\mathrm{(max)}\approx\frac{1}{2}$, $E_c/E_c^\mathrm{(max)}\approx 1$, $E_n/E_n^\mathrm{(max)}\approx 0$ and $E_n/E_n^\mathrm{(max)}\approx 1$ can still be distinguished in the realistic distributions and can thus serve as an experimental indication of the nearly fulfilled Efimov conditions.

The convergence for the realistic system is the hardest to achieve numerically -- one has to go up to $\rho_{max}=2.5a_{nn}$ and consequently include $N=12$ adiabatic channels to get the converged distributions. This calculation includes about 960 different hyper-spheric partial waves in the basis for the wave-function and seems to be converged within about 5\% accuracy.

\paragraph*{Conclusions.}

We have investigated the effects of the nearly fulfilled Efimov conditions on the properties of three-body resonances.  We have calculated the energy distributions of fragments in a three-body decay of a resonance for a realistic system, the ($1^-$) state in the halo nucleus $^{11}$Li within a three-body $^9{\rm Li}+n+n$ model, and two model systems where only $s$-wave interactions were allowed. We have shown that characteristic features appear as sharp peaks in the energy distributions and have traced their origin to the large scattering lengths between the decay fragments, like in the Efimov effect. 
The experimental
verification of the predicted distributions is straightforward but perhaps challenging due
to the required accuracy.

\paragraph*{Acknowledgements.} 

Continuous discussions with H. Fynbo and K. Riisager are highly
appreciated.

\end{document}